\documentclass[prb,twocolumn,superscriptaddress,showpacs]{revtex4-1}

\usepackage{amsmath}    
\usepackage{graphicx}  
\usepackage{bm}        
\usepackage{amssymb}   

\usepackage{color}

\newcommand{\bra}{\langle}
\newcommand{\ket}{\rangle}

\begin{document}

\title{Parameter-free hybrid-like functional based on an extended Hubbard model: DFT+$U+V$}%

\author{Nicolas Tancogne-Dejean}
  \email{nicolas.tancogne-dejean@mpsd.mpg.de}
 \affiliation{Max Planck Institute for the Structure and Dynamics of Matter and Center for Free-Electron Laser Science, Luruper Chaussee 149, 22761 Hamburg, Germany}
 \affiliation{European Theoretical Spectroscopy Facility (ETSF)}

 \author{Angel Rubio}
  \email{angel.rubio@mpsd.mpg.de}
\affiliation{Max Planck Institute for the Structure and Dynamics of Matter and Center for Free-Electron Laser Science, Luruper Chaussee 149, 22761 Hamburg, Germany}
 \affiliation{European Theoretical Spectroscopy Facility (ETSF)}
 \affiliation{Nano-Bio Spectroscopy Group, Universidad del Pa\'is Vasco, CFM CSIC-UPV/EHU-MPC, 20018 San Sebasti\'an, Spain }
\affiliation{Center for Computational Quantum Physics (CCQ), The Flatiron Institute, 162 Fifth Avenue, New York NY 10010}

\begin{abstract}
In this article, we propose an energy functional at the level of DFT+$U+V$ that allows us to compute self-consistently the values of the on-site interaction, Hubbard $U$ and Hund $J$, as well as the intersite interaction $V$. This functional extends the previously proposed ACBN0 functional [Phys. Rev. X 5, 011006 (2015)] including both on-site and intersite interactions. 
We show that this \textit{ab initio} self-consistent functional yields improved electronic properties for a wide range of materials, ranging from $sp$ materials to strongly-correlated materials. 
This functional can also be seen as an alternative general and systematic way to construct parameter-free hybrid functionals, based on the extended Hubbard model and a selected set of Coulomb integrals, and might be used to develop novel approximations. By extending the DFT+$U$ method to materials where strong local and nonlocal interactions are relevant, this work opens the door to the \textit{ab initio} study the electronic, ionic, and optical properties of a larger class of strongly correlated materials in and out of equilibrium.
\end{abstract}

\maketitle
\section{Introduction}

During the last few decades, density functional theory (DFT) has emerged as one of the most reliable and efficient numerical method to simulate a wide range of materials. However, it is well known that the most employed local and semilocal functionals suffer from many problems, in particular the so-called ``delocalization problem'', which prevents a practical application of using DFT to materials where strong local electron-electron are taking place\cite{anisimov1997first,himmetoglu2014hubbard}. More advanced functionals, such as meta-generalized gradient approximation, or hybrid functionals can solve some of these problems, but are still not ideal for strongly-correlated systems. In order to overcome this problem, one can rely on the well-established dynamical mean field theory (DMFT) ~\cite{PhysRevLett.62.324,PhysRevB.45.6479}, and its extensions such as DFT+DMFT or GW+DMFT, which has become the state-of-the-art methods to treat strongly correlated materials~\cite{RevModPhys.78.865}. In most of these methods, the effective electronic parameter describing the local interaction, the Hubbard $U$, is computed within the framework of constrained random-phase approximation (cRPA)~\cite{miyake_screened_2008,aichhorn_dynamical_2009, miyake_ab_2009,aichhorn_theoretical_2010}.

As an alternative effective approach, DFT+$U$ method originally proposed by V. Anisimov, A. Lichestein, and coworkers\cite{anisimov_band_1991,anisimov_density-functional_1993,liechtenstein_density-functional_1995,anisimov1997first} provides a successful way to improve the treatment of correlated solids upon DFT, without the numerical burden of the DFT+DMFT or GW+DMFT methods. 
In order to correct the over-delocalization of the electrons, it was proposed to include an energy penalty $U$ for the localized  $3d$ or $4f$ orbitals, in the spirit of the mean-field Hubbard model.~\cite{anisimov_band_1991,anisimov_density-functional_1993,liechtenstein_density-functional_1995,anisimov1997first,himmetoglu2014hubbard}
The success of the DFT+$U$ method mainly originates from the simplicity of the method, its relative low computational cost, and the fact that it can predict the proper magnetic ground state of  correlated materials such as charge-transfer and Mott insulators.\cite{anisimov1997first} 
Of course, the DFT$+U$ approach is not applicable to some really strongly-correlated systems and for these systems once needs to go beyond DFT$+U$ and use DMFT or cluster DMFT frameworks. Other limitations include for instance the incapacity to access lifetimes of quasiparticles. 
Still, the DFT+$U$ approach is very attractive when it comes to the calculation of larger systems, such as twisted bilayer systems with small twist angles~\cite{xian2019multiflat}, or for out-of-equilibrium situations, using real-time TDDFT$+U$ ~\cite{PhysRevLett.121.097402,topp2018all}. It also improves the description of the optical properties of some correlated materials within linear response~\cite{Implementation_DFTU}.  


Here, we develop an efficient numerical approach tailored towards materials for which not only local correlations are important, but also nonlocal correlations, which means a strong interaction between neighboring localized electrons. 
This is one step towards dealing with correlated materials in an efficient way but within the constraints of applicability of the DFT$+U+V$ method (as discussed for instance in Ref.~\onlinecite{himmetoglu2014hubbard}).
One example of such system in which nonlocal correlations are important is charge-ordering insulators, such as Fe$_3$O$_4$\cite{PhysRevB.54.4387}, for which an electron is delocalized over two sites and hence the Mott-Hubbard localization cannot occur. Nonlocal interaction also plays a key role in low-dimensional systems, such as ad-atoms on Si(111) surfaces~\cite{PhysRevLett.110.166401} or $sp$ electron systems such as graphite and graphene~\cite{tang2018role, PRL_intersite, PRL_intersite_old}. Finally, the role of the intersite interaction is strongly debated for high-$T_c$ superconductors and dictates many of their properties~\cite{PhysRevLett.58.2794}.
In all these systems, the intersite interaction plays a decisive role and the related self-interaction error contained in (semi-)local functional of DFT is crucially hampering the capability of these functionals.\\
The low-energy physics of these systems is usually well described by an extended Hubbard model. In particular, if we only account for charge interaction between neighboring sites, the extended Hubbard Hamiltonian reads
\begin{eqnarray}
 H_{\mathrm{eHub}} &=& -\sum_{i,j}\sum_\sigma t_{ij}(c_{i,\sigma}^\dag c_{j,\sigma} + \mathrm{h.c.})  + U\sum_i n_{i,\uparrow}n_{i,\downarrow} \nonumber\\
 &&+ \frac{1}{2}\sum_{i\neq j}\sum_{\sigma,\sigma'} V_{ij} n_{i,\sigma} n_{j,\sigma'},
\end{eqnarray}
where $\sigma$ denotes the spin index, $t_{ij}$ are the hopping matrix elements, $U$ represents the on-site interaction, and $V_{ij}$ are the nonlocal Coulomb matrix elements between the neighboring sites $i$ and $j$.

Our goal is to derive the expression of an orbital energy functional containing at the same time $U$ and $J$ describing the multi-band on-site interaction, and the intersite interaction $V$ describing the charge interaction between the different atomic sites. 
The functional can be seen as a hybrid functional, in which the Kohn-Sham orbitals are expanded into a basis of atomic-like orbitals, including the on-site terms, and some of the intersite terms. At variance with most of the proposed hybrid functionals, we do not use a mixing parameter to determine the weight of the exchange interaction, but we base our approach on the extended Hubbard model and the related DFT$+U+V$ scheme, which allows for a fully \textit{ab initio} and self-consistent estimate of the $U$, $J$, and $V$ effective electronic parameters. Moreover, we use an approximate double-counting term, as commonly done for DFT+$U$, which does not require any parameter to be adjusted.\\
 The self-consistency in the Hubbard $U$ can be crucial in the case of transition-metal complexes~\cite{PhysRevLett.97.103001}, and we expect this to be equally true for the intersite interaction. In our approach, the on-site Hubbard $U$, Hund $J$, and the intersite interaction $V$ are all evaluated at the same time, to ensure the consistency of our approach. It is important to stress that except for the additional cost of computing more Coulomb integrals at the beginning of the calculation, our approach does not represent a major extra cost compared to the usual DFT+$U$ and the ACBN0 functional\cite{Agapito_PRX,Implementation_DFTU}. 
This makes our method very attractive computationally. 
Another major interest of this  method is the possibility to  extend it directly to the time-dependent case, only assuming the adiabatic approximation, or to couple to other degrees of freedom, such as phonons.

This paper is organized as follow. First, we briefly review the DFT$+U$ and the DFT$+U+V$ methods in Sec.~\ref{sec:dftuv}. Then, we present our generalization of the ACBN0 functional, the extended ACBN0 functional, in Sec.~\ref{sec:ACBN0}. We then test our new functional on different systems and compare our results with prior works. Finally, we draw our conclusions in Sec.\ref{sec:conclusions}.

\section{DFT$+U$ and DFT$+U+V$}
\label{sec:dftuv}

The DFT$+U$ method aims at replacing the DFT energy functional $E_{\mathrm{DFT}}[n]$ by the DFT$+U$ energy functional of the form~\cite{Implementation_DFTU}
\begin{equation}
 E_{\mathrm{DFT+U}}[n,\{n_{mm'}^{I,\sigma}\}] = E_{\mathrm{DFT}}[n] + E_{ee}[\{n_{mm'}^{I,\sigma}\}] - E_{dc}[\{n_{mm'}^{I,\sigma}\}]\,,
 \label{eq:E_DFT_U}
\end{equation}
where $E_{ee}$ is the electron-electron interaction energy, and $E_{dc}$ accounts for the double counting of the electron-electron interaction already present in $E_\mathrm{DFT}$. 
This double-counting term is not known in the general case and several approximated forms have been proposed along the years. 
The $E_{ee}$ and $E_{dc}$  energies depend on the density matrix of a localized orbitals basis set $\{\phi^{I,n,l}_m\}$, which are the localized orbitals attached to the site $I$. In the following we refer to the elements of the density matrix of the localized basis as occupation matrices, and we denote them $\{n_{mm'}^{I,\sigma}\}$.
Combining these two expressions, we obtain the $E_U$ energy  to be added to the DFT total energy, which only depends on an effective  
Hubbard U parameter $U^{\mathrm{eff}}=U-J$. This gives the rotational-invariant form of DFT$+U$ proposed by Dudarev \textit{et al.},~\cite{PhysRevB.57.1505}
\begin{eqnarray}
 E_U[\{n_{mm'}^{I,\sigma}\}] = E_{ee}[\{n_{mm'}^{I,\sigma}\}] - E_{dc}[\{n_{mm'}^{I,\sigma}\}]\nonumber\\ 
  = \sum_{I,n,l} \frac{U^{\mathrm{eff}}_{I,n,l}}{2} \sum_{m,\sigma}\Big(n_{mm}^{I,n,l,\sigma} - \sum_{m'}n_{mm'}^{I,n,l,\sigma}n_{m'm}^{I,n,l,\sigma} \Big)\,,
  \label{eq:E_Dudarev}
\end{eqnarray}
where $I$ is an atom index, $n$, $l$ and $m$ refer to the principal, azimuthal, and angular quantum numbers, respectively, and $\sigma$ is the spin index.
In the case of a periodic system, the occupation matrices $n^{I,n,l,\sigma}_{mm'}$ are given by
\begin{equation}
n^{I,n,l,\sigma}_{mm'} = \sum_{n}\sum_{\mathbf{k}}^{\mathrm{BZ}} w_\mathbf{k}f_{n\mathbf{k}}^\sigma \bra\psi_{n,\mathbf{k}}^{\sigma} |\hat{P}^{I,n,l}_{mm'}|\psi_{n,\mathbf{k}}^{\sigma} \ket\,, 
\label{eq:occ_matrices}
\end{equation}
where $w_{\mathbf{k}}$ is the $\mathbf{k}$-point weight and $f_{n\mathbf{k}}^\sigma$ is the occupation of the Bloch state $|\psi_{n,\mathbf{k}}^{\sigma} \ket$.  
Here, $| \phi^{I,n,l}_{m}\ket$ are the localized orbitals that form the basis used to describe electron localization and $\hat{P}^{I,n,l}_{mm'}$ is the projector associated with these orbitals, usually defined as $\hat{P}^{I,n,l}_{mm'}=| \phi^{I,n,l}_{m}\ket  \bra \phi^{I,n,l}_{m'}|$. The definition of the orbitals and of the projectors will be discussed in more details below. In the following, we omit the principal quantum number $n$ for conciseness.

Recently, an extension of the DFT$+U$ method was proposed by V. Leiria Campo Jr and M. Cococcioni~\cite{0953-8984-22-5-055602}, in order to account for the intersite electronic interaction $V$, in the spirit of the extended Hubbard Hamiltonian~\cite{PhysRevLett.58.2794}. This represents the first \textit{ab initio} DFT$+U+V$ method proposed in the literature capable of estimating the intersite interaction $V$. In their method, only the charge interaction between neighboring sites is accounted for. In a similar spirit, Belozerov and coworkers proposed a LDA+DMFT+$V$ approach that they applied to the monoclinic phase of VO$_2$\cite{PhysRevB.85.045109}.\\
The Hubbard $U$ is usually defined as the averaged of the on-site interactions of the localized orbitals 
\begin{equation}
 U^{I,l} = \frac{1}{(2l+1)^2}\sum_{i,j}\bra \phi^{I,l}_{i}\phi^{I,l}_{j}|V_{ee}|\phi^{I,l}_{i}\phi^{I,l}_{j}\ket,
\end{equation}
where $V_{ee}$ is the screened Coulomb interaction. If the screened Coulomb interaction is frequency dependent, this leads to a frequency-dependent $U$, as used in the cRPA method. If a statically screened Coulomb interaction is used instead, as done in the DFT+$U$ method, the $U$ becomes frequency independent.\\ 
In a similar way, the most akin definition of the averaged intersite interaction $V$ is defined as~\cite{0953-8984-22-5-055602}
\begin{equation}
 V^{I,l}_{I',l'} = \frac{1}{(2l+1)(2l'+1)}\sum_{i,j}\bra \phi^{I,l}_{i}\phi^{I',l'}_{j}|V_{ee}|\phi^{I,l}_{i}\phi^{I',l'}_{j}\ket.
 \label{eq:averaged_V}
\end{equation}
For conciseness, we omit below the quantum numbers in our notation and refer in the following to $ V^{I,l}_{I',l'} $ as $V^{IJ}$. Similarly, we refer below to $U^{I,l}$ and $J^{I,l}$ as $U^I$ and $J^I$.
The expression for $E_{ee}$ and $E_{dc}$ become for DFT$+U+V$~\cite{0953-8984-22-5-055602}
\begin{eqnarray}
 E_{ee}[\{n_{mm'}^{IJ,\sigma}\}] &=& \sum_I\Big[\frac{U^I}{2} \sum_{m,m',\sigma}n_{mm}^{I,\sigma}n_{m'm'}^{I,-\sigma}\nonumber\\
 &&+\frac{U^I-J^I}{2} \sum_{m\neq m',\sigma}n_{mm}^{I,\sigma}n_{m'm'}^{I,\sigma}\Big]\nonumber\\
 &&+ \sum_{IJ}^{*} \frac{V^{IJ}}{2} \Big(N^{I}N^{J} -\sum_{m,m',\sigma}n^{IJ\sigma}_{mm'}n^{JI\sigma}_{m'm}\Big) \,, \nonumber\\
  E_{dc}[\{n_{mm'}^{IJ,\sigma}\}] &=& \sum_I\Big[\frac{U^I}{2}N^I(N^I-1) - \frac{J^I}{2}\sum_\sigma N^{I,\sigma}(N^{I,\sigma}-1)\Big] \nonumber\\
    &&+ \sum_{IJ}^{*} \frac{V^{IJ}}{2}N^IN^J\,,\nonumber
\end{eqnarray}
where $\sum_{IJ}^{*}$ denotes that for each atom $I$ the sum runs over its neighboring atoms $J$.
In these expression, we defined $N^{I,\sigma}=\sum_m n^{I,\sigma}_{mm}$ and $N^I = \sum_\sigma N^{I,\sigma}$.
This definition uses a generalization of the occupation matrix
\begin{equation}
 n^{IJ\sigma}_{mm'} = \sum_{n}\sum_{\mathbf{k}}^{\mathrm{BZ}} w_\mathbf{k}f_{n\mathbf{k}}^\sigma \bra\psi_{n,\mathbf{k}}^{\sigma} |\hat{P}^{IJ}_{mm'}|\psi_{n,\mathbf{k}}^{\sigma} \ket\,, 
\end{equation}
where it is clear that $n^{II\sigma}_{mm'}$ is the usual occupation matrix $n^{I\sigma}_{mm'}$ as defined in Eq.~\ref{eq:occ_matrices}.

Combining the previous expressions, one arrives to the energy $E_{UV}$ that must be added to the DFT total energy
\begin{eqnarray}
 E_{UV} &=& \sum_{I} \frac{U^{\mathrm{eff}}_{I}}{2} \sum_{m,\sigma}\Big(n_{mm}^{II,\sigma} - \sum_{m'}n_{mm'}^{II,\sigma}n_{m'm}^{II,\sigma} \Big) \nonumber\\
  &&- \sum_{IJ}^{*} \frac{V^{IJ}}{2} \sum_{m,m',\sigma}n_{mm'}^{IJ,\sigma}n_{m'm}^{JI,\sigma}.
  \label{eq:E_dudarev_UV}
\end{eqnarray}

This energy is the expression for the DFT$+U+V$ proposed in Ref.~\onlinecite{0953-8984-22-5-055602}, and is invariant under rotation of the orbitals of the same atomic site. This is a generalization of the work of Dudarev \textit{et al.}~\cite{PhysRevB.57.1505}, where the double-counting expression is a generalization of the fully-localized limit (FLL) double-counting of DFT$+U$. The motivation of this specific expression for the intersite interaction was done in Ref.~\onlinecite{0953-8984-22-5-055602} and is therefore not discussed here.
Below we show how it is possible to extend the work of Agapito \textit{et al.}~\cite{Agapito_PRX} to evaluate the average intersite interaction $V$ \textit{ab initio} and self-consistently, in a form of a pseudo-hybrid calculation.\\

%

\section{\textit{Ab initio} $V$:\\The extended ACBN0 functional}
\label{sec:ACBN0}

In Ref.~\onlinecite{Agapito_PRX}, an approximation to the electron interaction energy named ACBN0 functional was proposed, allowing for an efficient \textit{ab initio} evaluation of the DFT$+U$ energy, which can be seen as a screened Hartree-Fock evaluation of the on-site $U$, or equally as a pseudo-hybrid functional in which the (screened) Hartree-Fock energy is included only on a selected localized subspace. 
We propose here an extension of this approach, which includes not only the on-site interaction, but also the charge exchange between two sites. Below, we refer to this new functional as the extended ACBN0 functional.

In our generalized functional, the electron interaction energy is given
\begin{eqnarray}
E_{ee} = \frac{1}{2}\sum_I\Big[\sum_{\{m\}}\sum_{\alpha,\beta} \bar{n}_{mm'}^{I,\alpha}\bar{n}_{m''m'''}^{I,\beta} (\phi_m^I\phi_{m'}^I|\phi_{m''}^I\phi_{m'''}^I) \nonumber\\
- \sum_{\{m\}}\sum_\alpha \bar{n}_{mm'}^{I,\alpha}\bar{n}_{m''m'''}^{I,\alpha} (\phi_m^I\phi_{m'''}^I|\phi_{m''}^I\phi_{m'}^I)\Big] \nonumber\\
+ \frac{1}{4}\sum_{IJ}^*\Big[\sum_{mm'}\sum_{\alpha,\beta} \bar{n}_{mm}^{II,\alpha}\bar{n}_{m'm'}^{JJ,\beta} (\phi_m^I\phi_{m}^I|\phi_{m'}^J\phi_{m'}^J) \nonumber\\
- \sum_{mm'}\sum_\alpha \bar{n}_{mm'}^{IJ,\alpha}\bar{n}_{m'm}^{JI,\alpha}(\phi_m^I\phi_{m}^I|\phi_{m'}^J\phi_{m'}^J) \Big]\,,
\label{eq:HF_ACBN0_extended}
\end{eqnarray}
where we considered only the charge interaction between two sites, neglecting other two-site interactions and also three- and four-site interactions, as the former one is likely to be the largest contribution to the energy~\cite{0953-8984-22-5-055602}.

In Eq.~\eqref{eq:HF_ACBN0_extended}, the renormalized occupation matrices $\bar{n}^{I,\sigma}_{mm'}$ and occupations $\bar{N}^{I,\sigma}_{\psi_{n\mathbf{k}}}$ are respectively given by
\begin{equation}
\bar{n}^{I\sigma}_{mm'} = \sum_{n\mathbf{k}} w_\mathbf{k}f_{n\mathbf{k}} \bar{N}^{I\sigma}_{\psi_{n\mathbf{k}}}  \bra \psi_{n\mathbf{k}}^\sigma |\hat{P}^{I}_{mm'}| \psi_{n\mathbf{k}}^\sigma \ket \,,
\label{eq:notation_Pbar}
\end{equation}
  \begin{equation}
  \bar{N}^{I,\sigma}_{\psi_{n\mathbf{k}}} = \sum_{\{I\}} \sum_{m} \bra \psi_{n\mathbf{k}}^\sigma |\hat{P}^{I}_{mm}| \psi_{n\mathbf{k}}^\sigma \ket\,.
  \label{eq:notation_Nbar}
 \end{equation}
In the last expression, the sum $ \sum_{\{I\}} $ runs over all orbitals of the system owning the quantum numbers $n$ and $l$, and being attached to atoms of the same type as the atom $I$. This makes that $\bar{N}^{I,\sigma}_{\psi_{n\mathbf{k}}}$ can be seen as the Mulliken charge of atom $I$.~\cite{Agapito_PRX} 
Here we propose a definition of the generalized renormalized occupation matrices $\bar{n}^{IJ,\sigma}_{mm'}$
\begin{equation}
\bar{n}^{IJ\sigma}_{mm'} = \sum_{n\mathbf{k}} w_\mathbf{k}f_{n\mathbf{k}} \bar{N}^{IJ\sigma}_{\psi_{n\mathbf{k}}} \bra \psi_{n\mathbf{k}}^\sigma |\hat{P}^{IJ}_{mm'}| \psi_{n\mathbf{k}}^\sigma \ket \,,
\label{eq:notation_Pbar_IJ}
\end{equation}
where we introduced an ansatz similar to the one of the original paper of the ACBN0 functional, namely a degree of screening in $\bar{n}^{IJ,\sigma}_{mm'}$

\begin{equation}
 \bar{N}^{IJ\sigma}_{\psi_{n\mathbf{k}}} = \sum_{m} \bra \psi_{n\mathbf{k}}^\sigma |\hat{P}^{I}_{mm}| \psi_{n\mathbf{k}}^\sigma \ket\ + \sum_{m'} \bra \psi_{n\mathbf{k}}^\sigma |\hat{P}^{J}_{m'm'}| \psi_{n\mathbf{k}}^\sigma \ket\ \,,
\label{eq:notation_Nbar_IJ}
\end{equation}
where the quantity is not summed over all atomic species having the same quantum numbers (as done in Eq.~\ref{eq:notation_Nbar}), but corresponds here to the charge of the bound formed by atom $I$ and atom $J$.
Let us comment on the motivation for this renormalization of the generalized occupation matrix.
In the case of on-site interaction, the original motivation was that if a wavefunction is fully delocalized, i.e. not occupying the localized subspace, it should give a vanishing contribution to the on-site interaction $U$, and the ACBN0 functional should reduce to the (semi-)local functional used to describe the itinerant electrons. By including the weighting coefficient, which is the Mulliken charge of the set of orbitals, the effect is enhanced~\cite{Agapito_PRX}.
Here, as we consider an intersite interaction, the idea is the opposite. 
In the atomic limit, for which the wavefunctions are not delocalized over different sites, there should be no intersite interaction. 
This is well seen by the fact that for the Coulomb integral of the form $(\phi_m^I\phi_{m}^I|\phi_{m'}^J\phi_{m'}^J)$ to be non-zero, there must be an overlap of the charge density originating from the two atomic sites. Hence, if a wavefunction is not delocalized over the two considered sites, the term $\bar{N}^{IJ\sigma}_{\psi_{n\mathbf{k}}}$ vanishes, and the wavefunction does not contribute to the intersite $V$. Importantly, thanks to the renormalization employed here, if a wavefunction is fully localized on one site, the wavefunction still participates to the interaction. \footnote{We found that using $\sqrt{\bar{N}^{I\sigma}_{\psi_{n\mathbf{k}}}\bar{N}^{J\sigma}_{\psi_{n\mathbf{k}}}}$ as a definition of the weighting coefficient, which implies that a wavefunction only localized to one site does not contribute to $V$, overscreens the intersite interaction $V$, for instance for binary $s-p$ semiconductors. In this manuscript, all the results are presented for the weighting factor given by Eq.~\ref{eq:notation_Nbar_IJ}. } 
In the context of cRPA, the screening is given by the ``rest'' of the system, but also from the cross-term  screening coming from the ``rest'' of the system and from the localized subspace. Only the electrons fully localized in the localized subspace do not contribute to the screening. This means that $\bar{N}^{I\sigma}_{\psi_{n\mathbf{k}}}$ and  $\bar{N}^{IJ\sigma}_{\psi_{n\mathbf{k}}}$ should be equal to 1 for electrons fully contained in the localized subspace, as they do not lead to any screening. Partially delocalized electrons should also contribute to the screening. These properties are both given by the renormalization factor of the ACBN0 functional, as well as in our extended ACBN0 functional, which at least partly explains the success of this approach. 

\begin{widetext}
From Eq.~\eqref{eq:HF_ACBN0_extended}, the effective inter-site interaction between the I and J atomic sites, $\bar{V}^{IJ}$, is given by 
\begin{equation}
\bar{V}^{IJ} = \frac{1}{2}\frac{\sum_{mm'}\sum_{\alpha,\beta} (\phi_m^I\phi_{m}^I|\phi_{m'}^J\phi_{m'}^J)\Big[ \bar{n}_{mm}^{II,\alpha}\bar{n}_{m'm'}^{JJ,\beta} -\delta_{\alpha\beta} \bar{n}_{mm'}^{IJ,\alpha}\bar{n}_{m'm}^{JI,\alpha}\Big]}{\Big(\sum_{m,m',\alpha\beta}n^{I\alpha}_{mm}n^{J\beta}_{m'm'} -\sum_{m,m',\alpha}n^{IJ\alpha}_{mm'}n^{JI\alpha}_{m'm}\Big) }\,.
\label{eq:ACBN0_V}
\end{equation}
\end{widetext}

This expression is the main result of this section. 
With it, one can evaluate the intersite interaction \textit{ab initio} and self-consistently, similarly to what is done for the effective $U$ in the ACBN0 functional. We also derived the expression of the extended ACBN0 functional for the case of noncollinear spins, as presented in Appendix~\ref{app_dftuv_nc}. The expression of the forces is also given in Appendix~\ref{app_forces}.

%

So far, we remained elusive on the orbitals used to define the localized subspace. This is a technical but very important implementation-dependent issue that deserves some discussion.
In the Octopus code~\cite{Implementation_DFTU,Octopus_paper_2019}, we construct the localized orbitals $\{ \phi^{I,n,l}_{m} \}$ by taking the radial part of the pseudo-atomic wavefunctions given by the pseudopotential files, and multiplying them by the usual spherical harmonics, in order to obtain the pseudo-atomic orbitals. More precisely, in case of periodic solids, we use in all the above equations not the isolated localized orbital, but the Bloch sums of the localized orbitals, which read as
\begin{equation}
  \phi^{I,n,l}_{m,\mathbf{k}}(\mathbf{r}) = \frac{1}{\sqrt{N}}\sum_{\mathbf{R}} e^{-i\mathbf{k}\ldotp\mathbf{R}}\phi^{I,n,l}_{m}(\mathbf{r}+\mathbf{R})\,,
  \label{eq:orbitals}
\end{equation}
and which amount for introducing a phase factor when the spheres on which the atomic orbitals are defined cross the border of the real-space simulation box. Here $N$  corresponds to the number of unit cells forming the periodic crystal, i.e., the number of $\mathbf{k}$-points of the simulation.  The projection onto Kohn-Sham states selects a single momentum, explaining why the $\mathbf{k}$-point index was not specified in the above equations. Also note that the normalization factor in the Bloch sum vanishes when we use the periodicity of the crystal for computing the sum over the entire crystal. This reduces to a sum over the unit cell without the normalization factor.

In order to be able to treat various type of solids, including weakly correlated solids such as Si, we also implemented the L\"owdin orthonormalization procedure, which transforms the set of non-orthogonal localized orbitals $\{ \phi^{I,n,l}_{m} \}$ into an orthonormal set of localized orbitals
\begin{equation}
\bar{\phi}_{i,\mathbf{k}}(\mathbf{r}) = \sum_{j} \left(S^{-\frac{1}{2}}_{\mathbf{k}}\right)_{ji} \phi_{j,\mathbf{k}}(\mathbf{r})\,,
\end{equation}
where $i$ and $j$ indices run over all the considered orbitals, and the overlap matrix for the set of considered orbitals is $(S_\mathbf{k})_{ij} = \bra \phi_{i\mathbf{k}}| \phi_{j\mathbf{k}}\ket$. Importantly, using these orthogonalized orbitals, we obtain a trace of the on-site occupation matrix which is consistent with the Mulliken population analysis and leads to the exact same trace of the occupation matrix than the dual projector defined in Ref.~\onlinecite{PhysRevB.73.045110} for non-orthogonal basis set.

\begin{figure}[t]
  \begin{center}
.    \includegraphics[width=0.8\columnwidth]{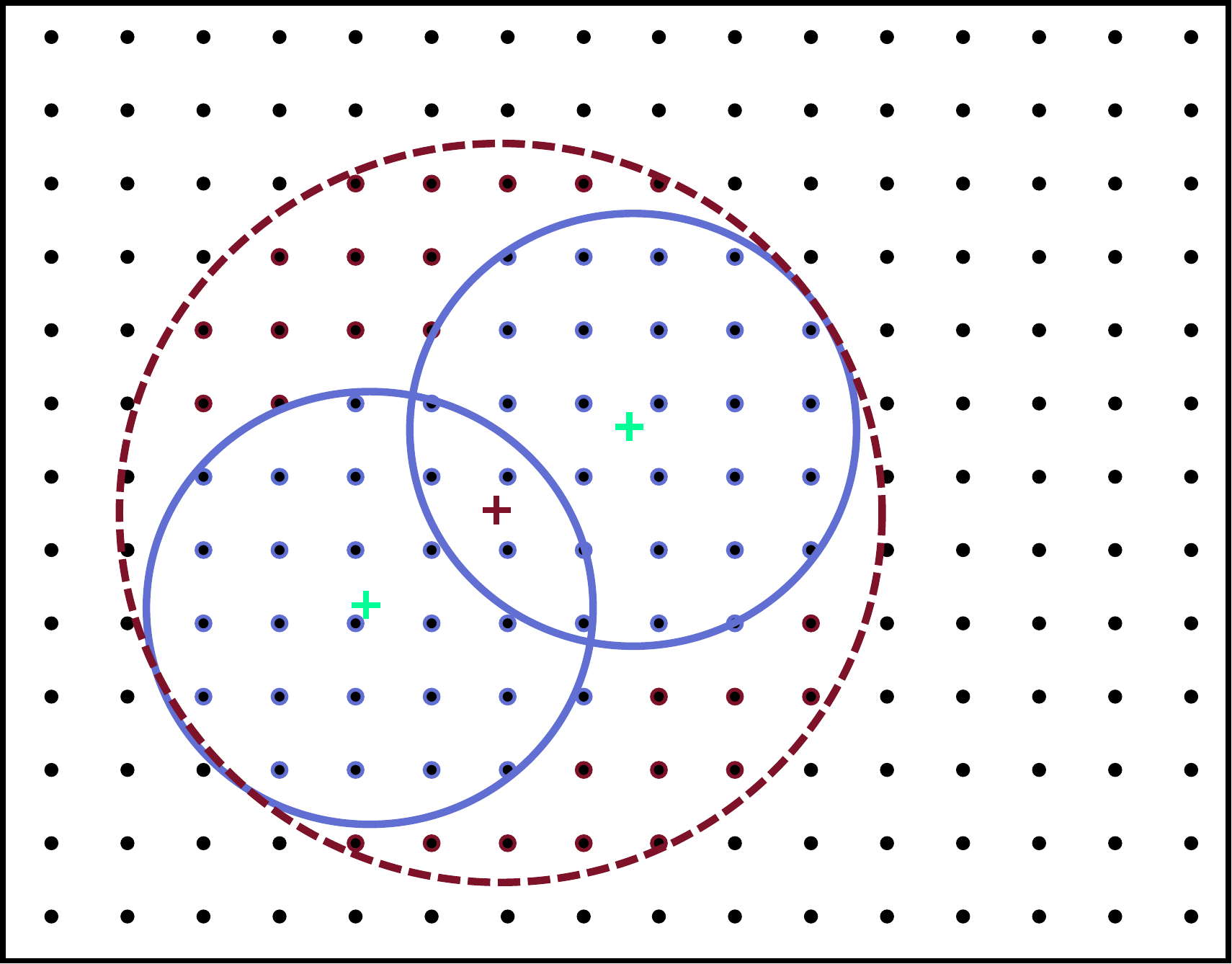}
    \caption{\label{fig:Sketch_orbs} Sketch of possible choices for defining a portion of the real-space grid on which the atomic orbitals and Coulomb integrals can be evaluated. Atom positions are indicated by green crosses. Details are discussed in the main text.
    }
  \end{center}
\end{figure}

Due to the periodicity of the Bloch sums of the localized orbitals, we only need to compute projections on the orbitals of the atoms inside the simulation box, irrespective of the number of neighboring atoms considered. This is also the case for a simpler DFT$+U$ calculation, making the cost of DFT$+U+V$ calculation only mildly more expensive than a more standard DFT$+U$ calculation. This is a major difference compared to the original formulation of DFT+$U+V$, which requires the construction of larger supercells to include further neighbors~\cite{0953-8984-22-5-055602}.\\
%
%
We stress here that the orthonormal orbitals given by Eq.~\eqref{eq:orbitals} are not the ones that we use to compute the Coulomb integrals, as they are periodic orbitals, and Coulomb integrals computed using these orbitals would contain both on-site and intersite overlaps, which is not what we want.
For this reason, the Coulomb integrals are computed before performing the orthonormalization procedure, from the atomic orbitals of the pseudopotential.
More precisely, each of them is evaluated on a portion of the grid, and the Coulomb integrals are computed on the union of these two spheres, using a non-periodic Poisson solver.
This is sketched in Fig.~\ref{fig:Sketch_orbs}, in which the violet points correspond to the grid points obtained from the union of the two spherical meshes centered on two atoms (indicated by green crosses). These are the points we used in our implementation. Another possible choice would be a single, larger, sphere centered at the middle of the two atoms, as indicated in red in Fig.~\ref{fig:Sketch_orbs}. This choice obviously leads to more grid points and we found that there is no difference 
 in between the two choices, as the atomic orbitals rapidly decay away from the center of the atom.
 Finally, we note that a formulation in terms of Wannier functions would look very much the same as the one we have presented here.

\section{Results}
\label{sec:results}
%
\subsection{ZrSiSe}
\begin{figure}[t]
  \begin{center}
.    \includegraphics[width=0.9\columnwidth]{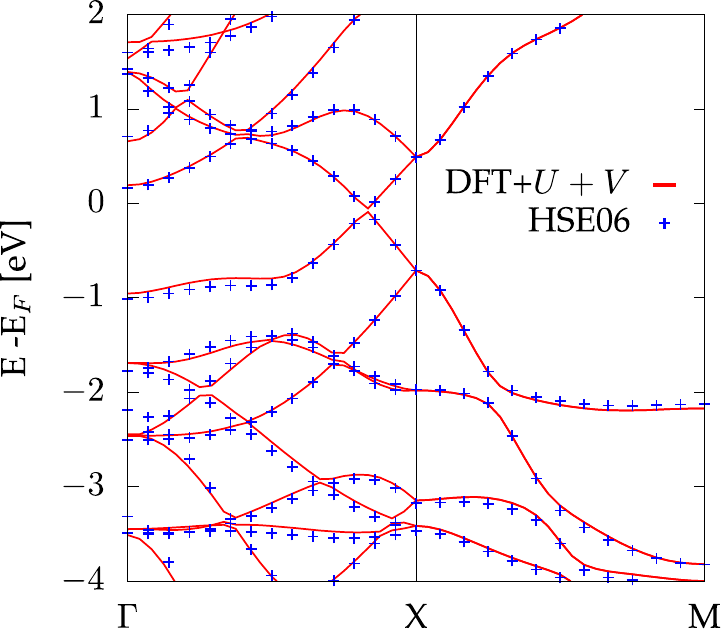}
    \caption{\label{fig:ZrSiSe} Bandstructure of bulk ZrSiSe obtained using the extended ACBN0 functional (red lines) compared to the hybrid functional calculation (blue dots). }
  \end{center}
\end{figure}

The nodal-line semimetals have received a lot of attention recently. Due to the vanishing density of states at the Dirac or Weyl points, the screening of the Coulomb interaction is altered, and long-range Coulomb interaction is a crucial ingredient in the description of these materials~\cite{PhysRevB.93.035138}. Hence, the nodal-line semimetals exhibit strong nonlocal correlations~\cite{PhysRevB.97.075140,PhysRevLett.120.216401}, which are not captured by a local Hubbard $U$ as used in DFT+$U$. This represents therefore an interesting potential application of our extended ACBN0 functional. \\
In order to benchmark our functional, we decided to investigate ZrSiSe, and we used a hybrid functional as a reference to compare with. In Ref.~\onlinecite{PhysRevB.97.075140,ZrSiSe}, it was shown that hybrid functional calculations with a fraction of  exact exchange of 7\% reproduce best the experimental results, compared to the standard fraction of 25\% used in the HSE06 functional\cite{heyd2003hybrid}. 
We employed the experimental lattice constant of 3.623 $\AA$ and we sampled the real-space using a spacing of 0.3 Bohr and the Brillouin zone using a $7\times7\times3$ $\mathbf{k}$-point grid. We considered localization of the $d$ orbitals of Zr and included the interaction with the first nearest neighbors. We employed the local-density approximation for the DFT exchange-correlation energy functional. We obtained self-consistent values for the effective $U_{\mathrm{eff}}=U-J$ and $V$ of 1.63\,eV and 0.37\,eV respectively.
The comparison of the bandstructure computed from DFT+$U+V$ and the hybrid functional is shown in Fig.~\ref{fig:ZrSiSe}. We found that DFT+$U+V$ has almost the same band dispersion than the one obtained from the hybrid functional close to the Weyl point, showing the validity of our functional in describing this material. \\
We recently applied our functional to successfully reproduce the measured angle-resolved photoemission spectrum of ZrSiSe~\cite{ZrSiSe} including spin-orbit coupling. Formulae for noncollinear spins are presented in Appendix~\ref{app_dftuv_nc}.

\subsection{Graphene and graphite}
Low-dimensional $sp$ materials received a lot of attention recently as they display strong local and nonlocal Coulomb interaction\cite{PRL_intersite,PRL_intersite_old}.
In Ref.~\onlinecite{PRL_intersite}, cRPA calculations of the on-site and intersite interactions for graphene at half-filling were presented. Similar calculations were performed in graphene and graphite in Ref.~\onlinecite{PRL_intersite_old}.
In order to illustrate the flexibility and robustness of our implementation, we compute the intersite interactions for graphene, treated here with mixed periodic boundary conditions, and graphite, which is a fully periodic material. We use a $15\times15$ $\mathbf{k}$-point grid to sample the two-dimensional Brillouin zone of graphene and a $12\times12\times4$ grid in the case of graphite. We employ an in-plane lattice constant of 2.47$\AA$ in both cases, and an out-of-plane constant of 6.708$\AA$ for graphite. The real-space grid is sampled by a spacing of 0.45 Bohr and we employed the norm-conserving Pseudodojo pseudopotential~\cite{van2018pseudodojo}.
We compare the previously reported values to the ones obtained by our functional in Tab.~\ref{tab:intersite_comp}. A major difference is that the in the present calculations all $p$ orbitals are considered, whereas prior studies only considered $p_z$ orbitals. As a result, both on-site and intersite interactions are found to be smaller in our case that in previous works\cite{PRL_intersite,PRL_intersite_old}, which is fully compatible with considering more orbitals in the localized subspace, which, in the language of cRPA, reduces the screening from the rest of the system.

\begin{table}[h]
\begin{ruledtabular}
\begin{tabular}{l|c|c|c|c|c}
Material & \multicolumn{2}{c|}{Graphite} & \multicolumn{3}{c}{Graphene} \\
         & This work & Ref.~\onlinecite{PRL_intersite_old} & This work & Ref.~\onlinecite{PRL_intersite_old} & Ref.~\onlinecite{PRL_intersite}   \\
         \hline
U     & 7.62& 8.0-8.1& 7.58 & 9.3 & 10.16  \\
$V_{01}$ & 4.04 &  3.9 & 4.00 & 5.5 & 5.68 \\
$V_{02}$ & 2.58& 2.4-2.4 & 2.56 & 4.1 & 4.06 \\
$V_{03}$ & 2.25 & 1.9 & 2.22& 3.6 & 3.70 \\
$V_{04}$ & & - & 1.88 & - & 3.19 \\
\end{tabular}
\end{ruledtabular}
\caption{\label{tab:intersite_comp}Calculated values of the on-site ($U$) and intersite ($V_{0i}$) interactions for graphite and graphene computed with our extended ACBN0 functional (DFT+U+V), compared to prior works. Values are given in eV. The notation $V_{0i}$ denotes the intersite interaction between an atom in the unit cell, and its $i$-th neighbor. In the case of graphite, two values are indicated corresponding to the two sublattices of the system.}
\end{table}

Already in the 1950s, in the context of $\pi$-conjugated systems, it was proposed by Pariser, Parr, and Pople a one-band model with nearest-neighbor hopping and intersite interaction. This model has been widely studied and few expressions have been proposed to interpolate the Coulomb interaction between the long-range $1/r$ behavior, and the short-range on-site value, which is the Hubbard $U$.
In order to get a more physical insight on the values obtained by our method, we compare them to the popular Ohno interpolation formula~\cite{ohno1967molecular}, which reads
\begin{equation}
 V_{ij} = \frac{U}{\sqrt{1+(U\epsilon r_{ij})^2}}.
\end{equation}
In this expression, the intersite interaction between atoms $i$ and $j$ separated by the distance $r_{ij}$ is estimated from the on-site interaction $U$ and an effective dielectric constant $\epsilon$. 
\begin{figure}[t]
  \begin{center}
.    \includegraphics[width=0.9\columnwidth]{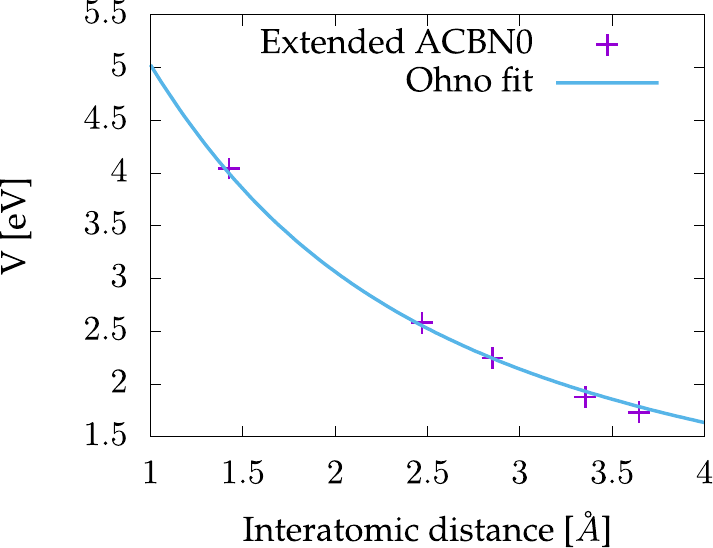}
    \caption{\label{fig:Ohno} Comparison between the calculated intersite $V$ for graphite and the extrapolated values using the Ohno formula and the calculated on-site effective $U$. An effective dielectric constant $\epsilon$ of 2.15 is used here. }
  \end{center}
\end{figure}

The fact that we can fit the values of the intersite interaction with the Ohno potential indicates that the Coulomb integrals are properly computed. The most interesting point is that the effective dielectric constant $\epsilon=2.15$, used here to match our calculated intersite interaction values, agrees reasonably very well with the effective dielectric constant of graphite of 2.5 found experimentally or from cRPA calculations ~\cite{PRL_intersite_old}. This shows that the functional correctly describes the screening at place in graphite.

\begin{figure}[t]
  \begin{center}
.    \includegraphics[width=\columnwidth]{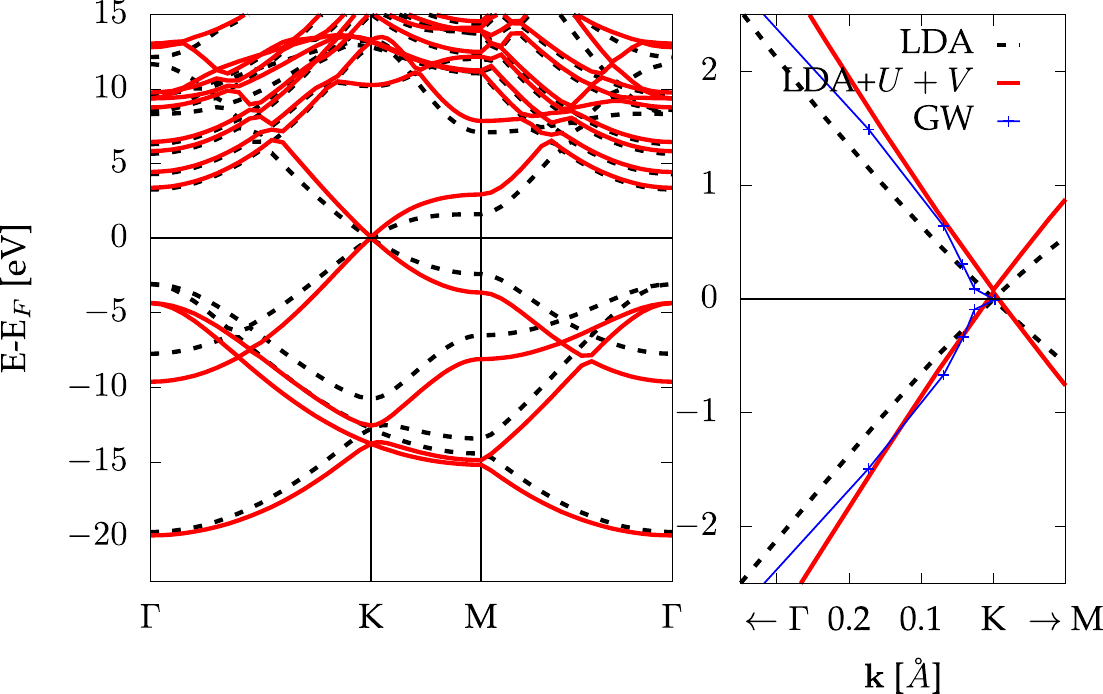}
    \caption{\label{fig:graphene} Left panel: Band structure of graphene, calculated using the LDA (dashed black lines), compared to the LDA$+U+V$ calculation (red lines). Right-panel: comparison of the band-dispersion close to the $K$ point. The GW data (blue dotted lines) are taken from Ref.~\onlinecite{trevisanutto2008ab}. }
  \end{center}
\end{figure}

Fig.~\ref{fig:graphene} shows the band structure of graphene obtained using the extended ACBN0 functional, compared to the LDA one. In this case, we included both $s$ and $p$ orbitals, orthonormalized using the L\"owdin orthonormalization procedure discussed above. The comparison of the band dispersion close to the $K$ point (right panel of Fig.~\ref{fig:graphene}) shows that the extended ACBN0 leads to Fermi velocity quite close to the GW one around the $K$ point, compared to the LDA calculation. This demonstrates that our functional also improves the description of the electronic properties of graphene.
However, we note that the extended ACBN0 functional opens too much the gap between the $\pi$ bands at the $M$ point. Indeed, the extended ACBN0 functional yields a splitting of 6.58\,eV, compared to 4.01\,eV for the LDA. This overestimates the GW splitting of 4.89\,eV, but is quite similar to the B3LYP value of 6.14\,eV ~\cite{PhysRevB.78.081406}, which is probably linked to the way we are approximating the screening in the ACBN0 functional.

\subsection{Silicon}

\begin{figure}[t]
  \begin{center}
.    \includegraphics[width=0.9\columnwidth]{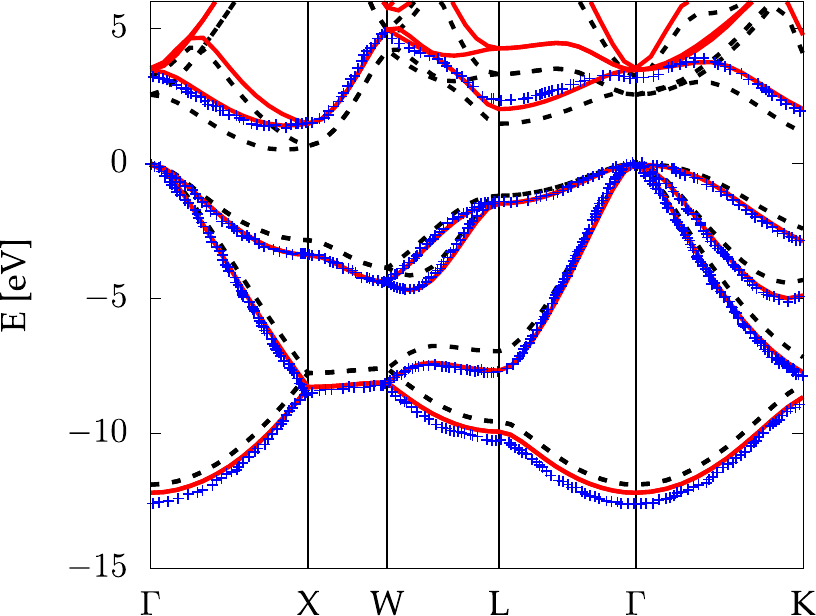}
    \caption{\label{fig:BS_Si} Band structure of bulk silicon calculated using the LDA (dashed black lines) , compared to the one obtained from the LDA$+U+V$ method (red lines). The blue dots correspond to the LDA$+U+V$ calculation from linear-response calculation of $U$ and $V$ in supercell~\cite{0953-8984-22-5-055602}. Only valence bands and the first conduction band are shown.
    }
  \end{center}
\end{figure}

In Ref.~\onlinecite{0953-8984-22-5-055602}, LDA$+U+V$ method was applied to bulk silicon, as a test case of weakly correlated semiconductor material that is well understood. They showed that this corrective functional improves the values of direct and indirect bandgaps compared to LDA.
The calculated values of on-site $U$ and intersite $V$ interactions, obtained from our functional, are compared in Tab.~\ref{tab:intersite_comp_Si} to the values obtained from linear-response calculation in $3\times3\times3$ supercell calculation~\cite{0953-8984-22-5-055602}. 
We employed an $8\times8\times8$ $\mathbf{k}$-point grid to sample the Brillouin zone, a grid spacing of 0.5 Bohr and norm-conserving Pseudodojo pseudopotential~\cite{van2018pseudodojo}. The lattice constant is taken to be the experimental one of $a=5.431\AA$.
Overall, our results are found to be in reasonable agreement with the values of Ref.~\onlinecite{0953-8984-22-5-055602}, even if we are getting significant differences for some of the values. This can be tracked to different implementations, different pseudopotentials, and the difference between the supercell treatment and the primitive one, as the screening is computed in a different way in the two approaches. However, it is worth noting that the band structure of silicon computed from LDA+$U+V$ almost matches perfectly the one obtained from linear response values, as shown in Fig.~\ref{fig:BS_Si}. This shows that whereas the values of $U$ and $V$ are not fully transferable from one implementation to another one, the observables obtained from the two approaches, the extended ACBN0 functional and linear response, are very similar. We checked that using the generalized gradient approximation instead of the LDA one for the exchange-correlation part does not lead to a significant change of the calculated values of the effective electronic parameters.
 
\begin{table}[h]
\begin{ruledtabular}
\begin{tabular}{l|c|c}
Parameters & This work & Ref.~\onlinecite{0953-8984-22-5-055602}  \\
         \hline
$U_{ss}$     & 3.68 & 2.82 \\
$U_{pp}$     & 3.55 & 3.65 \\
$U_{sp}$     & 2.29 & 3.18 \\
$V_{ss}$     & 0.94 & 1.40 \\
$V_{sp}$     & 1.37 & 1.36 \\
$V_{pp}$     & 1.86 & 1.34 \\
\end{tabular}
\end{ruledtabular}
\caption{\label{tab:intersite_comp_Si} Effective electronic parameters of silicon calculated from the extended ACBN0 functional, compared to the one obtained by linear response~\cite{0953-8984-22-5-055602}. The parameters labeled by $U$ correspond to on-site interactions whereas the ones labeled by $V$ correspond to the nearest neighbor interactions. }
\end{table}

\subsection{Transition metal oxides: MnO and NiO}

\begin{figure}[t]
  \begin{center}
    \includegraphics[width=0.9\columnwidth]{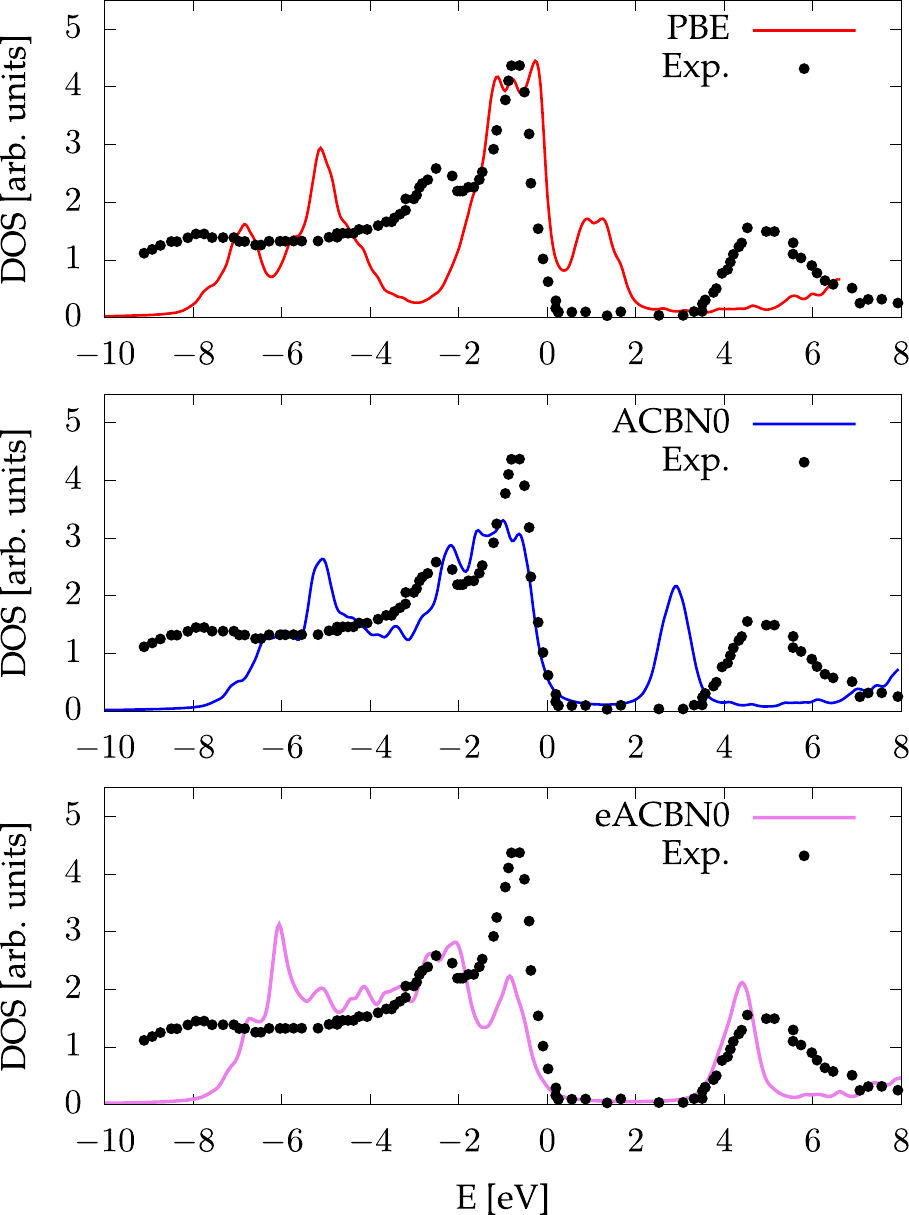}
    \caption{\label{fig:DOS_NiO} Comparison of the density of states of NiO obtained using the LDA (top panel),the ACBN0 functional (DFT+$U$; middle panel), and the extended ACBN0 functional (DFT+$U+V$; bottom panel), with the experimental one obtained from photoemission and inverse photoemission\cite{PhysRevLett.53.2339}. The energies were shifted such that the top of the valence band corresponds to the zero of energy in all cases.
    }
  \end{center}
\end{figure}

\begin{figure}[t]
  \begin{center}
    \includegraphics[width=0.9\columnwidth]{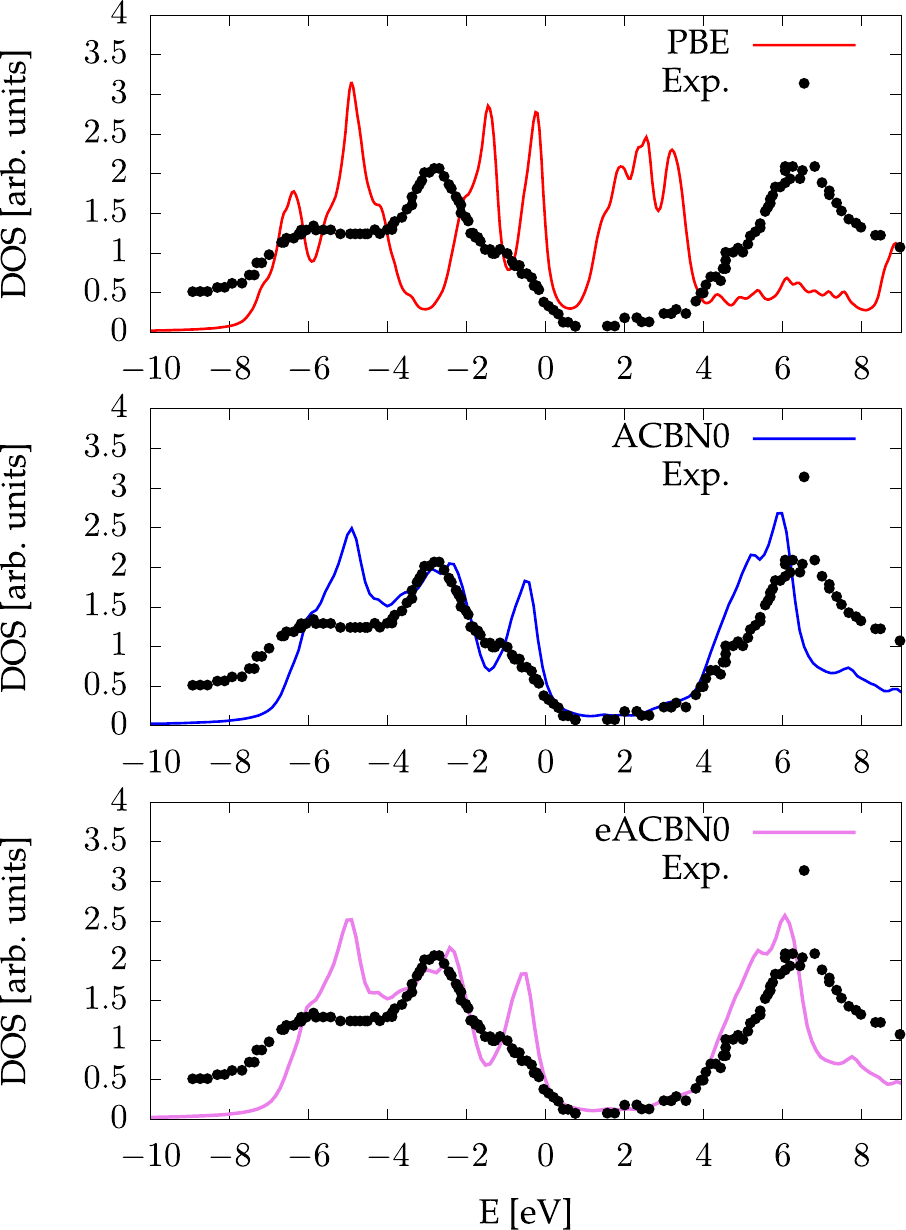}
    \caption{\label{fig:DOS_MnO} Same as Fig.~\ref{fig:DOS_NiO}, but for bulk MnO. Experimental data are taken from Ref.~\onlinecite{PhysRevB.44.1530}.
    }
  \end{center}
\end{figure}

We now turn our attention to transition metal oxides, for which it is quite common to employ the DFT+$U$ method to improve the description of electronic correlations.
We computed the density of state of both bulk NiO and bulk MnO in their antiferromagnetic phase, as shown in Figs.~\ref{fig:DOS_NiO} and ~\ref{fig:DOS_MnO}. The magnetic moment obtained for the different functionals are reported in Tab.~\ref{tab:mag_mom}. 
All the calculations presented here were performed considering type-II antiferromagnetic materials below their his N\'eel temperature.
Below its N\'eel temperature  ($T_N=523$K\cite{cracknell_space_1969}), NiO exhibits a rhombohedral structure, which is obtained by contraction of the original cubic cell along one of the [111] directions.\cite{cracknell_space_1969} 
However, we have neglected the small distortions and considered NiO (and MnO) in its cubic rock-salt structure. Calculations were performed using a lattice parameter of 4.1704~\AA\ for NiO and 4.4315~\AA\ for MnO, a real-space spacing of $\Delta \mathbf{r}=0.2$ Bohr, and a $8\times8\times8$ $\mathbf{k}$-point grid to sample their Brillouin zones. We employed norm-conserving pseudo-potentials, and all orbitals available were considered for localization. A broadening of 0.1\,eV was used to mimic the experimental broadening of the density of states.

\begin{table}[h]
\begin{ruledtabular}
\begin{tabular}{l|c|c}
Functional & NiO & MnO  \\
         \hline
PBE  & 1.15 & 4.13 \\
ACBN0  & 1.57 & 4.39 \\
eACBN0   & 1.64 & 4.36 \\
Exp. & 1.64\footnote{Reference ~\onlinecite{Alperin}}, 1.90\footnote{Reference ~\onlinecite{PhysRevB.27.6964}} & 4.58\footnote{Reference ~\onlinecite{PhysRevB.27.6964}}, 4.79\footnote{Reference ~\onlinecite{doi:10.1063/1.1668855}}\\
\end{tabular}
\end{ruledtabular}
\caption{\label{tab:mag_mom} Calculated magnetic moments of the transition metal atoms for different methods, in Bohr magneton, compared to the experimental values. The magnetic moments are computed by taking a sphere around the atoms, with respectively a radius of 1.11\AA\ and 1.04\AA\ for NiO and MnO. }
\end{table}

As many prior studies, see for instance Refs.~\onlinecite{mandal2019systematic,PhysRevB.74.155108}, we found that the semi-local Perdew-Burke-Ernzerhof (PBE)~\cite{PhysRevLett.77.3865} functional produces a very small bandgap for NiO.
Adding a Hubbard $U$ improves the energy bandgap compared to the experiment. This also changes the shape of the peaks located around -5eV and -7eV. Considering the intersite interaction (bottom panel of Fig.~\ref{fig:DOS_NiO}) improves further the bandgap. Moreover, it induces a splitting of the broad peak obtained at the PBE+$U$ level into two peaks, corresponding to O-$p$ bands and Ni-$d$ bands, and improves the position of the satellite peak to be at -6 eV.
These three peaks make that the density of states computed at the extended ACBN0 level is improved compared to the calculation considering only the on-site interaction, even if notable differences compared to the experiment still persists. It is important to note that these effects are the same as the one obtained from self-consistent linear-response evaluation of the on-site and intersite interactions, which shows again that our approach qualitatively captures the same effects as the linear-response method of Ref.~\cite{0953-8984-22-5-055602}. We find that the magnetic moment of NiO is also improved using the extended ACBN0 functional compared to the PBE+$U$ result, see Tab.~\ref{tab:mag_mom}.

The calculation performed for bulk MnO also show, at the PBE level, the same effects as reported before in the literature\cite{mandal2019systematic,PhysRevB.74.155108}: there is a wrong splitting of the density of states for the top valence bands, the bandgap is too small, and the bottom of the valence bands shows a double peak, as for NiO, which is also not correct. We found that these features are already corrected very well by the addition on an on-site interaction, which yield a nice agreement with the experimental result.
Interestingly, the addition of an intersite interaction does not produces any sizable difference, compared to a simple PBE+$U$ calculation. The magnetic moment of the Mn atom is also not much affected by the inclusion of the intersite interaction, see Tab.~\ref{tab:mag_mom}.

These calculations show that in the case where the on-site interaction is not enough to describe the electronic properties, the \textit{ab initio} evaluation of the intersite interaction improves the description of the electronic properties, whereas when the on-site interaction is enough, adding the intersite interaction does not degrade the electronic properties.

\subsection{Band gaps of semiconductors and insulators}

After investigating in details the effect of our functional on the electronic properties of specific materials, we finally want to discuss its performance on the electronic band gaps of a series of semiconductors and insulators. It is indeed interesting to wonder if this functional can be used without having to rely on some knowledge of how the localization takes place, and to use it as a pseudo-hybrid functional.
Our results, presented in Tab.~\ref{tab:bandgaps}, are obtained using a $8\times8\times8$ $\mathbf{k}$-point grid for sampling the Brillouin zone of each material. We employed the experimental lattice constant and used PBE norm-conserving Pseudodojo pseudopotential~\cite{van2018pseudodojo}.
For each materials we converged the spacing such that it yields a converged PBE bandgap at 20\,meV.
In the case of ACBN0 and eACBN0 calculations, all atomic orbitals available from the pseudopotentials were employed to construct the localized subspaces.
This implies that different pseudopotentials will lead to different intersite interactions, as for instance some pseudopotentials provide $d$ orbitals for Si, whereas some other do not. Overall, we find that the choice of the pseudopotential does not affect too much the band gaps obtained by the functional, except in the case of partially occupied $d$ orbitals, such as in transition metal oxides, and there we note that some important deviations can be observed, as for example in NiO. 

\begin{table}[h]
\begin{ruledtabular}
\begin{tabular}{lccccc}
 & PBE & ACBN0 & eACBN0 & $G_0W_0$ & exp. \\
         \hline
Ar & 8.66 & 13.08  & 13.07 & 13.28\cite{PhysRevLett.102.226401} & 14.20\cite{PhysRevLett.102.226401} \\        
C  & 4.19 & 4.28 & 5.71 & 5.50\cite{PhysRevLett.102.226401} & 5.48\cite{PhysRevLett.102.226401}\\   
Si & 0.61 & 0.44 & 1.30 & 1.12\cite{PhysRevLett.102.226401} & 1.17\cite{PhysRevLett.102.226401} \\   
Ge & 0.04 & 0.00 &  0.03 & 0.66\cite{PhysRevLett.102.226401} & 0.74\cite{PhysRevLett.102.226401} \\   
LiF &9.14 & 11.13 & 12.19 & 13.27\cite{PhysRevLett.102.226401} & 14.20\cite{PhysRevLett.102.226401}\\
MgO&4.78 & 5.54 & 6.71 & 7.25\cite{PhysRevLett.102.226401} & 7.83\cite{PhysRevLett.102.226401}\\
BN& 4.30& 5.30& 5.78 & 5.85\cite{PhysRevLett.102.226401} & 6.25\cite{PhysRevLett.102.226401}\\
AlSb& 1.22& 1.14 & 1.78 & 1.80\cite{PhysRevMaterials.3.064603} & 1.62\cite{PhysRevMaterials.3.064603}\\
AlP& 1.56& 1.77 & 2.83 & 2.39\cite{PhysRevMaterials.3.064603} & 2.45\cite{PhysRevLett.102.226401}\\
AlAs& 1.43&  1.46 & 2.47 & 2.09\cite{PhysRevMaterials.3.064603}& 2.15\cite{PhysRevMaterials.3.064603} \\
GaAs& 0.52& 0.35 & 1.14 & 1.64\cite{PhysRevLett.102.226401} & 1.52\cite{PhysRevLett.102.226401}\\
NiO& 1.01& 2.14 & 2.29& 1.1\cite{PhysRevLett.102.226401}& 4.0~\cite{Agapito_PRX} \\
MnO& 0.99& 0.24& 0.32& 1.7\cite{PhysRevLett.102.226401} & 4.1~\cite{Agapito_PRX}\\
TiO$_2$&1.91 & 2.61 & 2.83 & 3.18\cite{PhysRevLett.102.226401}  & 3.03-3.3~\cite{Agapito_PRX}\\
ZnO& 0.95& 3.51  & 3.61 & 2.68\cite{PhysRevLett.102.226401} & 3.44~\cite{Agapito_PRX}\\
\hline
MARE(\%) & 48.9 & 38.97 & 19.48 & 14.71&
\end{tabular}
\end{ruledtabular}
\caption{\label{tab:bandgaps} Electronic band gaps, in eV, calculated as the differences of Kohn-Sham eigenvalues for various materials. Calculations were performed at the experimental lattice constant. The mean absolute relative error (MARE) is also reported.}
\end{table}

The mean absolute relative error obtained for the eACBN0 functional (19.48\%) shows that overall the functional improves drastically the energy bandgap of semiconductor and insulators, compared to PBE (which yields a MARE of 48.9\%) and PBE+$U$, as this latter only mildly improves upon PBE. The results obtained with the eACBN0 are reasonably close to the one obtained with a by-far more expansive $G_0W_0$ method, which gives for this set of materials a MARE of 14.71\%.\\
This shows that beyond the set of specific materials presented above, the functional properly improves the description of electronic properties for a large variety of solids, without having to select \textit{a priori} on which type of orbitals the localization has to take place.


%
%
%
%

%
%
\section{Conclusion}
\label{sec:conclusions}
%
In conclusion, we presented an efficient method to compute \textit{ab initio} and self-consistently the effective electronic parameters $U$, $J$, and $V$. We implemented the DFT$+U+V$ and our novel energy functional
in the real-space TDDFT code Octopus\cite{Octopus_paper_2019}.
We showed results for ground-state calculations showing that our implementation yields results in good agreement with the ones previously reported in the literature.
%
We applied our functional to a correlated nodal-line semimetal, ZrSiSe, showing that our functional produces very similar results to the one obtained from a by far more expensive hybrid functional calculation. 
Applied to low-dimensional $sp$ compounds, our functional gives results in qualitative agreement with cRPA calculations. We tested our functional on bulk silicon and bulk transition metal oxides, and we found that our functional reproduces well the results of linear-response in supercell~\cite{0953-8984-22-5-055602}.
Finally, we applied it to a set of semiconductors and insulators, showing that the electronic bandgap is well improved compared PBE and PBE+$U$, and that the functional performs reasonably well compared to $G_0W_0$.\\
Let us comment on the choice of localized orbitals. In this work we employed pseudo-atomic orbitals obtained from the pseudopotentials. Whereas we found that for $s-p$ semiconductors, the choice of the pseudopotential is quite irrelevant for the obtained results, we observed significant differences for transition metal oxides depending on the choice of the pseudopotentials. This can be seen as a limitation of our method. However, our approach is not limited to pseudopotential-based codes and can straightforwardly be used using any type of localized orbitals, such as for instance Wannier orbitals. One can therefore get rid of the pseudopotential dependency, or at least reduce it drastically, by applying the present method on Wannier orbitals constructed on-the-fly, such as for instance using the SCDM-k method\cite{DAMLE20171}\\
Finally, we note that in this work we followed Ref.~\onlinecite{0953-8984-22-5-055602} and only considered specific intersite interaction. Determining how reliable and general is this approximation will require further investigations and would require extending the presented energy functional to include other intersite interactions. The method we presented here is general enough such that one could easily extend it to include other interaction terms. The present work also suggests that when evaluating the expensive exchange operator, most of the Coulomb integrals might not be very relevant and some of them can maybe be neglected, based on physical considerations.\\
The extension of this functional to the time-dependent case, or its performance on forces and vibrational properties of solids, will be investigated in a future work.

\acknowledgments
N.T.-D. would like to acknowledge M. A. Sentef for interesting and fruitful discussions.
This work was supported by the European Research Council (ERC-2015-AdG694097), the Cluster of Excellence ‘Advanced Imaging of Matter' (AIM), Grupos Consolidados (IT1249-19) and SFB925. The Flatiron Institute is a division of the Simons Foundation.

\appendix
\section{DFT+U+V with noncollinear spin}
\label{app_dftuv_nc}

In this section we present the formula we obtained as we extended the DFT+$U$+$V$ to the noncollinear spin case.
Based on the same approximation as before, we arrive to the following expression for the electron-electron interaction energy for noncollinear spin systems
\begin{eqnarray}
  E_{ee} &=& \sum_{I}\frac{U^I}{2} \sum_{\sigma}\sum_{mm'} n_{mm}^{\sigma\sigma}n_{m'm'}^{-\sigma-\sigma}  + \frac{U^I-J^I}{2} \sum_{\sigma}\sum_{m\neq m'} n_{mm}^{\sigma\sigma}n_{m'm'}^{\sigma\sigma}\nonumber\\
  &&- \frac{U^I}{2} \sum_{\sigma}\sum_{m} n_{mm}^{\sigma-\sigma}n_{mm}^{-\sigma\sigma}
  - \frac{J^I}{2} \sum_{\sigma}\sum_{m\neq m'} n_{mm}^{\sigma-\sigma}n_{m'm'}^{-\sigma\sigma}
  \nonumber\\
&& + \sum_{I\neq J}\frac{V_{IJ}}{2} \sum_{\sigma\sigma'} \sum_{mm'}\Bigg[n_{mm}^{I\sigma\sigma} n_{m'm'}^{J\sigma'\sigma'} -n_{mm'}^{IJ\sigma\sigma'} n_{m'm}^{JI\sigma'\sigma}\Bigg]\,.
\label{eq:ee_SOC}
\end{eqnarray}

In the fully localized limit, the corresponding double counting term is given for the on-site interaction by~\cite{PhysRevB.80.035121} 
\begin{equation}
 E_{DC}^{\mathrm{on-site}} = \sum_{I}\Bigg[\frac{U^I}{2}N^I(N^I-1) - \frac{J^I}{2}N^I(\frac{N^I}{2}-1) - \frac{J^I}{4}\mathbf{m}^I\ldotp\mathbf{m}^I\Bigg]\,,
\end{equation}
where $\mathbf{m}$ is the magnetization of the localized subspace~\cite{Implementation_DFTU}, and $N^I = \sum_{\sigma}\sum_m n_{mm}^{I,\sigma\sigma}$ is the number of electrons in the localized orbitals of the site $I$. 

For the inter-site interaction, the double counting term is given by 
\begin{equation}
 E_{DC}^{\mathrm{inter-site}} = \sum_{IJ}^*\frac{V}{2}N^IN^{J}\,.
\end{equation}

Putting everything together, one obtains that the rotationally-invariant form corresponding to Eq.~\eqref{eq:E_dudarev_UV} for the noncollinear spins reads
\begin{eqnarray}
 E_{UV} &=&  \sum_I\frac{U_{I}-J_{I}}{2}  \left[ \sum_{\sigma}\sum_{m}n_{mm}^{I,\sigma\sigma} - \sum_{mm'}\sum_{\sigma\sigma'}n_{mm'}^{I,\sigma\sigma'}n_{m'm}^{I,\sigma'\sigma}\right]\nonumber\\
 &&-\sum_{IJ}^*\frac{V_{IJ}}{2}\sum_{mm'}\sum_{\sigma\sigma'}n_{mm'}^{IJ,\sigma\sigma'}n_{m'm}^{JI,\sigma'\sigma} \,.
\end{eqnarray}
which is very similar to the expression Eq.~\eqref{eq:E_dudarev_UV} presented in the main text, with the exception that the trace is done also on the spin coordinates for the on-site and inter-site interactions.
The corresponding potential is given by
\begin{eqnarray}
 V_{UV}|\psi_{n\mathbf{k}}^\sigma\ket &=&  \sum_I\frac{U_{I}-J_{I}}{2}  \sum_{mm'}\sum_{\sigma'}\left[\delta_{mm'}\delta_{\sigma\sigma'} - 2n_{m'm}^{I,\sigma'\sigma}\right]P_{mm'}^{II}|\psi_{n\mathbf{k}}^{\sigma'}\ket\nonumber\\
 &&-\sum_{IJ}^*V_{IJ}\sum_{mm'}\sum_{\sigma\sigma'}e^{i\mathbf{k}\ldotp\mathbf{R_{IJ}}}n_{m'm}^{JI,\sigma'\sigma} P_{mm'}^{IJ}|\psi_{n\mathbf{k}}^{\sigma'}\ket\,.
\end{eqnarray}

\begin{widetext}
Based on the expression Eq.~\ref{eq:ee_SOC}, we obtain the expression for the inter-site $V$ for the non-colinear spin case as
\begin{equation}
\bar{V}^{IJ} = \frac{1}{2}\frac{\sum_{mm'}\sum_{\sigma\sigma'} (\phi_m^I\phi_{m}^I|\phi_{m'}^J\phi_{m'}^J)\Big[ \bar{n}_{mm}^{I,\sigma\sigma}\bar{n}_{m'm'}^{J,\sigma'\sigma'} -\bar{n}_{mm'}^{IJ,\sigma\sigma'}\bar{n}_{m'm}^{JI,\sigma'\sigma}\Big]}{\sum_{m,m'}\sum_{\sigma\sigma'}\Big(n^{I\sigma\sigma}_{mm}n^{J\sigma'\sigma'}_{m'm'} -n^{IJ\sigma\sigma'}_{mm'}n^{JI\sigma'\sigma}_{m'm}\Big) }\,.
\label{eq:ACBN0_V_SOC}
\end{equation}
\end{widetext}

\section{Forces}
\label{app_forces}
%

In this section we present the formula we obtained for the contribution to the forces acting on an atom $\alpha$ along the direction $i$ coming from the intersite term. This is given by
\begin{eqnarray}
 F_{\alpha,i}^V = -\frac{\partial E_V}{\partial R_{\alpha,i}} = -\frac{\partial E_V}{\partial n_{mm'}^{IJ,\sigma}}\frac{\partial n_{mm'}^{IJ,\sigma}}{\partial R_{\alpha,i}}\nonumber\\
 =  \sum_{IJ}^{*} \frac{V^{IJ}}{2} \sum_{m,m',\sigma}\Bigg[ n_{m'm}^{JI,\sigma}\frac{\partial n_{mm'}^{IJ,\sigma}}{\partial R_{\alpha,i}} +  n_{mm'}^{IJ,\sigma}\frac{\partial n_{m'm}^{JI,\sigma}}{\partial R_{\alpha,i}} 
  \Bigg].
\end{eqnarray}
As for the case of DFT+$U$~\cite{Implementation_DFTU}, the derivative of the generalized occupation matrix is expressed in terms of the derivative of the orbitals to reduce the so-called egg-box effect. After some algebra, one obtains that
\begin{eqnarray}
\frac{\partial n_{m'm}^{JI,\sigma}}{\partial R_{\alpha,i}}  = \sum_{\mathbf{k},v}w_{\mathbf{k}}f_{\mathbf{k},v} \Bigg[   \delta_{\alpha,J}\bra \phi_{m',\mathbf{k}}^{J,n,l}|\frac{\partial \psi_{\mathbf{k},v}^\sigma}{\partial r_{i}}\ket\bra \psi_{\mathbf{k},v}^\sigma|\phi_{m,\mathbf{k}}^{I,n,l}\ket 
\nonumber\\
+
   \delta_{\alpha,I}\bra \phi_{m',\mathbf{k}}^{J,n,l}|\psi_{\mathbf{k},v}^\sigma\ket \bra\frac{\partial\psi_{\mathbf{k},v}^\sigma}{\partial r_{i}}| \phi_{m,\mathbf{k}}^{I,n,l}\ket\Bigg]\,.\nonumber\\
\end{eqnarray}

In the present expression, as in the ACBN0 case~\cite{Implementation_DFTU} the derivative of $U$, $J$, and $V$ to the forces is not taken into account. Moreover, in case of a L\"owdin orthonormalization, the contribution from the derivative of the overlap matrix with respect to the atomic position is missing in the previous expression. This will be explored in a further publication.

\bibliography{bibliography}

\end{document}